# Orbital torque switching of perpendicular magnetization in light metal/ferrimagnet bilayers

Teng Xu[1,*], Aihua Tang[1], Kang Wang[1], Yizhou Liu[1,2], and Haifeng Du[1,2,*]

[1]*Anhui Province Key Laboratory of Low-Energy Quantum Materials and Devices, High Magnetic Field Laboratory, HFIPS, Chinese Academy of Sciences, Hefei 230031, China*

[2]*Science Island Branch of Graduate School, University of Science and Technology of China, Hefei 230026, China*

[*]Correspondence should be addressed to:

txu@hmfl.ac.cn

duhf@hmfl.ac.cn

## Abstract

**Orbital torque, associated with orbital current, enables light metals to efficiently manipulate magnetization with rich tunability. A clear demonstration of perpendicular magnetization switching using light metals alone is essential for understanding orbital physics and developing high-density orbitronic devices. Here, we report orbital torque switching of perpendicular magnetization in light metal (Ti, V, Cr)/ferrimagnet ($Fe_{1-x}Gd_x$) bilayers. Taking the Ti/ $Fe_{1-x}Gd_x$ sample as a model system, the torque efficiency increases four-fold by enhancing the spin-orbit coupling in $Fe_{1-x}Gd_x$ through modulating Gd composition, which is a characteristic feature of orbital torque. Our findings demonstrate that light metals in combination with rare earth-transition metal ferrimagnets can be employed for efficient orbitronic devices and serve as a model system for studying orbitronics.**

**Introduction**

Achieving effective control of magnetization through electrical means is central to modern information technology [1-3]. Methods harnessing the electron's spin angular momentum, such as the spin-orbit torque (SOT), have already facilitated vast device applications [4-7]. Recently, the electron's orbit degree of freedom has sparked attention as an emergent way to manipulate magnetization [8-10], known as orbital torque (OT). OT is associated with orbital current generated by either interfacial or bulk effect, e.g., the orbital Hall effect (OHE). In the case of OHE, an external electric field induces a transverse orbital current [11-16], which can then be converted into spin angular momentum and exert an OT on the magnetization [17-24] (Fig. 1a). The generation of orbital current does not require spin-orbit interaction (SOC), so even light metals (LMs), such as Ti, can be utilized as sources of OT [25,26].

OT has already been characterized in various LM-magnetic bilayers/multilayers [20,21,23,27-29]. It has been shown that the torque efficiency of LMs is comparable or even greater than that of heavy metals (HMs) usually employed for SOT [11,12,15,30]. The correlation between OT and the adjacent magnetic layer's SOC also provides a new avenue for enhancing the torque efficiency [17,18]. Therefore, LMs can control magnetization with the advantages of being environmentally friendly and low-cost due to the diverse material choices.

To realize high-density orbitronic applications with OT and further unlock its technological potential, it is crucial to demonstrate the switching of perpendicular magnetization with LMs [9,10]. Previous efforts have employed LM/HM/ferromagnet multilayers and Zr-magnetic bilayers/multilayers, where large torque efficiency has been identified [21,22,31,32]. However, the sign changes of the measured torque and the multilayer structure make it difficult to distinguish the contributions of OT and SOT as well as the bulk and interfacial effects. The role of SOC from the magnet side has also not been well explored in the switching. Hence, a clear demonstration of an OT switching of perpendicular magnetization via LMs alone remains elusive.

Here, we report on OT in light metal-ferrimagnet (Ti, V, Cr/Fe$_{1-x}$Gd$_x$) bilayers that are large enough to induce full switching of the perpendicular magnetization. The OT efficiency, extracted via harmonic voltage measurement, shows that it is dominated by the bulk orbital Hall effect (OHE). Furthermore, the rare-earth transition metal (RE-TM) ferrimagnet employed here enables additional control of its SOC while maintaining almost identical magnetic parameters. Taking advantage of this feature, we found that the OT efficiency in Ti/Fe$_{1-x}$Gd$_x$ shows a four-fold increase when the ferrimagnetic layer's SOC is enhanced, which is a key characteristic of OT. Our results show that LMs alone can switch perpendicular magnetization with great tunability from the magnet side.

**Perpendicular magnetization switching with Ti**

To demonstrate the OT-induced perpendicular magnetization switching, a typical light metal Ti is employed. Ti has negligible spin-orbit coupling so it shows negligible spin Hall conductivity but rather large orbital Hall conductivity [11,12,15,23-25]. Ti/ferrimagnet bilayers with stacking order Ti(10)/Fe$_{0.70}$Gd$_{0.30}$(10)/Si$_3$N$_4$ (number denotes the thickness in nanometer) were fabricated on thermally oxidized silicon substrates by using ultrahigh vacuum magnetron sputtering system. The Fe$_{1-x}$Gd$_x$ alloy is a typical RE-TM ferrimagnet with two inequivalent and antiparallel magnetic sublattices. A finite net magnetization remains because of the uncompensated magnetic moments from the oppositely aligned sublattices [33-35]. The magnetic properties of Fe$_{1-x}$Gd$_x$ ferrimagnets are highly tunable through the chemical compositions. The Fe$_{0.70}$Gd$_{0.30}$ composition is magnetically Gd-dominant in which the Gd sublattice is parallel with the net magnetization. To study the OT switching, the samples were patterned into Hall bar structures as shown in Fig. 1b, which also illustrates the measurement configuration. The out-of-plane magnetic hysteresis loop (Fig. 1c, Extended Data Fig. 1 and Supplementary Fig. S1) and the anomalous Hall effect (AHE) loop (Fig. 1d) indicate that the ferrimagnet possesses strong perpendicular magnetic anisotropy.

Next, we examine the OT-induced magnetization switching in Ti(10)/Fe$_{0.70}$Gd$_{0.30}$ device. Due to the strong SOC of Gd, the orbital current injected from the Ti layer can be converted into spin current, which further applies a torque on the local magnetic moments of the ferrimagnet through exchange coupling. During the current-induced orbital torque switching measurements, a series of current pulses with varied amplitude and fixed duration were applied along longitudinal direction (the $x$ axis) of Hall bar devices. Simultaneously, a longitudinal in-plane magnetic field $H_x$ was applied along the current pulse direction, which is required for realizing a deterministic switching. In between two adjacent current pulses, the magnetization states of the $Fe_{0.70}Gd_{0.30}$ can be inferred from the AHE resistance ($R_{xy}$). Note that the current density in Ti layers ($J_C$) can be estimated through the parallel resistor model and more details can be found in Extended Data Fig. 2 and Supplementary Note 2.

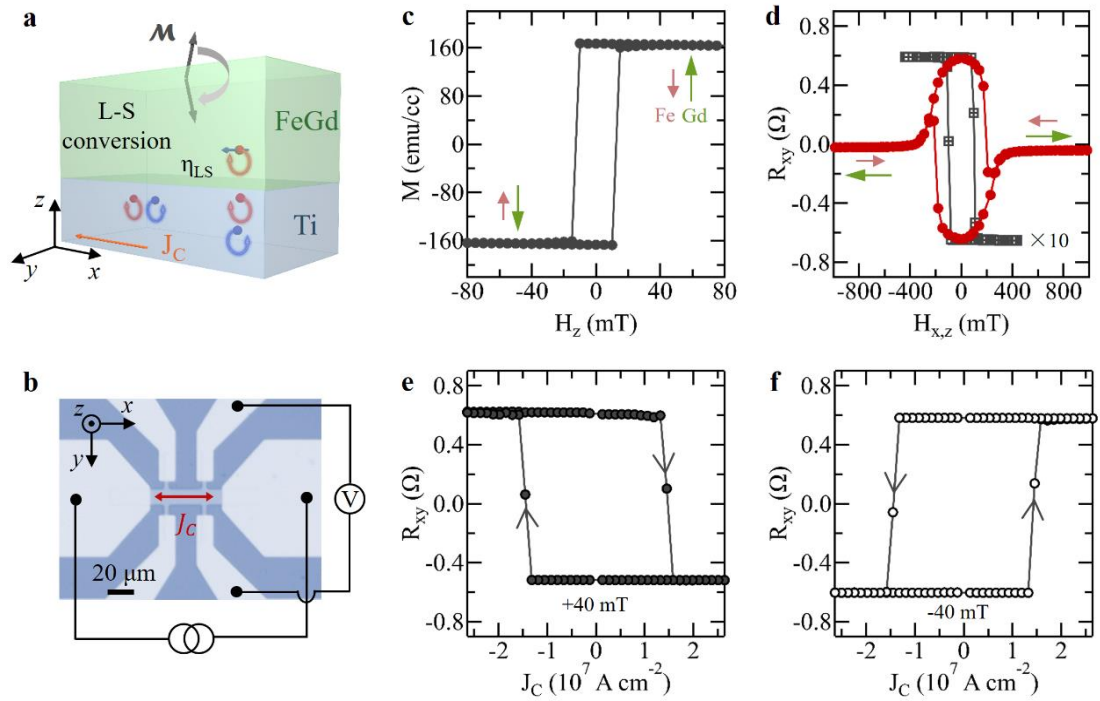

**Fig. 1. Perpendicular magnetization switching with Ti. a,** Schematic of the orbital torque switching of perpendicular magnetization in Ti/ferrimagnet bilayer. The orbital current is generated through the orbital Hall effect in Ti, which is then injected into the adjacent ferrimagnet. The orbital-to-spin current conversion is directly achieved due to the moderate spin-orbit coupling of Gd. Lastly, the converted spin current interacts with the local magnetic moment through the exchange coupling and gives rise to the orbital torque. $\eta_{LS}$ denotes the orbital-to-spin current conversion efficiency in ferrimagnet. **b,** Optical image of the Hall bar device and the corresponding measurement configuration. **c,** The out-of-plane magnetic hysteresis loop for the Ti(10)/Fe$_{0.70}$Gd$_{0.30}$(10) sample. **d,** The corresponding anomalous Hall (black) and planar Hall (red) loops. The pink (green) arrow denotes the spin orientation of Fe (Gd) magnetic sublattice. **e, f,** The current-induced magnetization switching curves under an in-plane field $H_x = +40$ mT and $H_x = -40$ mT. The arrows denote

the switching polarities.

Figs. 1e and f show the AHE resistance as a function of the current pulse with varying amplitudes under an in-plane field $H_x = \pm 40\ mT$. Compared with the field-driven AHE resistance change ($\Delta R_{xy}$), a nearly complete current-induced switching is achieved. The critical switching current density $J_C^{th}$ is $\sim 1.5 \times 10^7\ A\ cm^{-2}$. The value is comparable to those originally reported for the SOT-driven switching in HM/ferromagnet devices [36-38]. Fig. 2a further illustrates the current-induced magnetization switching cycles in the Ti(10)/Fe$_{0.70}$Gd$_{0.30}$ device, indicating the high repeatability and stability of the OT switching. The current-induced switching reverses its polarity after applying opposite in-plane field $H_x$, which is consistent with the symmetry of OT. Additionally, the critical switching current density reduces with increased $H_x$ (Extended Data Fig. 3 and Supplementary Fig. S2), which occurs because of the reduced energy barrier.

To verify the crucial role of Ti, control film stack of Si$_3$N$_4$/Fe$_{0.70}$Gd$_{0.30}$(10)/Ti(10) and Si$_3$N$_4$/Fe$_{0.70}$Gd$_{0.30}$(10)/Si$_3$N$_4$ and the corresponding devices were also fabricated and examined. For the inverted film stack, the OT should reverse its sign due to the opposite polarization of orbital current generated in Ti layer and so does the switching polarity. As shown in Fig. 2b, the experimental switching polarity reverses the sign compared with the original film stack as expected. Additionally, no magnetization switching can be observed in single Fe$_{0.70}$Gd$_{0.30}$(10) layer in the absence of Ti. Furthermore, the energy dispersive spectroscopy (EDS) element mapping demonstrates that there is no obvious variation in the chemical composition along the thickness direction of FeGd layer (Extended Data Fig. 4). Hence, the recently discovered self-torque in RE-TM ferrimagnets [39-42] can also be excluded and the current-induced switching observed here is attributed to the OT originating from Ti. Additionally, to confirm the universal OT in LM/FeGd bilayer, we have further performed OT switching in V/Fe$_{0.80}$Gd$_{0.20}$/Si$_3$N$_4$ and Cr/Fe$_{0.80}$Gd$_{0.20}$/Si$_3$N$_4$ samples. OT-induced deterministic full switching are demonstrated as shown in Extended Data Fig. 5.

**Thickness dependence of the orbit torque switching**

To further characterize the OT, current-induced switching measurements were performed in other Ti($t_{Ti}$)/Fe$_{0.70}$Gd$_{0.30}$ devices with varying the Ti thickness. As shown in Fig. 2c, deterministic OT switching can be achieved under an in-plane external field $H_x = -40\ mT$. Their corresponding switching curves with sign reversal at $H_x = +40\ mT$ can be found in the Extended Data Fig. 6. The critical switching current density ($J_C^{th}$) reduces to $\sim 1.0 \times 10^7\ A\ cm^{-2}$ with the Ti thickness up to 12 nm, as illustrated in Fig. 2d.

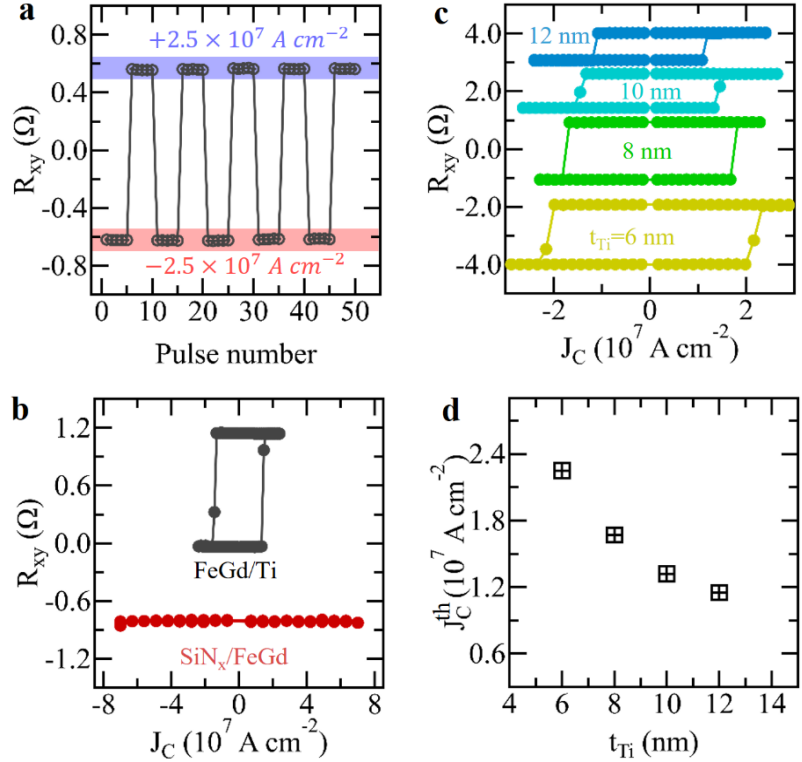

**Fig. 2. Thickness dependence of the orbit torque switching. a**, Repeatable magnetization switching by a series of current pulses of $\pm 2.5 \times 10^7$ A cm$^{-2}$ at $H_x = -40$ mT. **b**, The current-induced magnetization switching curves for the inverted Si$_3$N$_4$/Fe$_{0.70}$Gd$_{0.30}$(10)/Ti(10) (grey) and Si$_3$N$_4$/Fe$_{0.70}$Gd$_{0.30}$(10)/Si$_3$N$_4$ (red) stacking. **c**, Current-induced magnetization switching curves for Ti($t_{Ti}$)/Fe$_{0.70}$Gd$_{0.30}$(10)/Si$_3$N$_4$ multilayers with varying Ti thickness at $H_x = -40$ mT. **d**, The evolution of the critical switching current density $J_C^{th}$ as a function of Ti thickness.

To quantify the OT efficiency, the harmonic Hall voltage measurements were performed using a low-current excitation lock-in technique. We applied a low-frequency ac current to the Hall bar and simultaneously measure the first harmonic ($V_{1\omega}$) and second harmonic ($V_{2\omega}$) Hall voltage with two lock-in amplifiers by sweeping the in-plane field $H_x$. When a sinusoidal current is applied, the current-induced effective field oscillates in sync with the applied current. Based on the macrospin model, the oscillating effective fields induce small oscillation of the magnetization about its equilibrium direction, generating a second harmonic contribution to the Hall voltage [43-46]. The first harmonic Hall term ($V_{1\omega}$) relates to the equilibrium direction of the magnetization and is independent of the effective fields. The second harmonic term ($V_{2\omega}$) measures the susceptibility of the magnetization to the current-induced effective fields. In general, the effective filed can be decomposed into the damping-like field $H_{DL}$ and field-like field $H_{FL}$. When the in-plane field $H_x$ is larger than the anisotropy field $H_k$, the magnetization is nearly aligned along the external field and the secondary harmonic Hall voltage signals can be written as [47-50]

$$V_{2\omega} = \frac{V_{AHE}}{2}\frac{H_{DL}}{|H_x - H_k|} + V_{PHE}\frac{H_{FL}}{|H_x|} + V_{thermal}\frac{H_x}{|H_x|} \qquad (1)$$

where the $V_{AHE}$ and $V_{PHE}$ are the anomalous Hall and planar Hall voltage coefficients and $V_{thermal}$ is the thermal contribution from the anomalous Nernst and spin Seebeck effects. Note that the thermal contributions can be measured and separated by studying the field dependence of $V_{2\omega}$ in the high field limit, which can be found in Supplementary Fig. S3. In addition, since the $V_{PHE}$ is much smaller than $V_{AHE}$ for the perpendicular FeGd ferrimagnets, the second term can be neglected. Thus, the effective damping-like field $H_{DL}$ is mostly relevant to the current-induced magnetization switching.

Figs. 3a and b show the first and second harmonic Hall voltage ($V_{1\omega}$ and $V_{2\omega}$) of Ti(10)/Fe$_{0.70}$Gd$_{0.30}$(10)/Si$_3$N$_4$ sample as a function of $H_x$ at different current densities, respectively. As shown in Fig. 3a, the net magnetization is aligned along the in-plane direction when $H_x$ is larger than $H_k$. Through fitting the $V_{2\omega} - H_x$ curves with the equation (1), the effective damping-like field ($H_{DL}$) and magnetic anisotropy field ($H_k$) can be acquired. The first and second harmonic Hall signals for other samples of various Ti thickness can be found in Supplementary Figs. S4-S8. The variations of the $H_{DL}$ as a function of the current density in all samples are summarized in Fig. 3c. By a linear fitting, the OT efficiency ($\chi_{OT} = H_{DL}/J_C$) can be determined. Fig. 3d shows the evolution of orbital torque efficiency with the Ti thickness. As Ti thickness increases, the orbital torque efficiency first increases and then saturates at $t_{Ti} \approx 10$ nm. The increase of $\chi_{OT}$ with increasing the Ti thickness suggests that the OT is dominated by the bulk effect rather than the interfacial effect such as the orbital Rashba-Edelstein effect [51-54]. When the Ti thickness is 12 nm, the OT efficiency reaches $1.3 \times 10^{-7}\ mT cm^2 A^{-1}$. This increasing OT efficiency is consistent with the decreasing critical switching current density with Ti thickness shown in Fig. 2d.

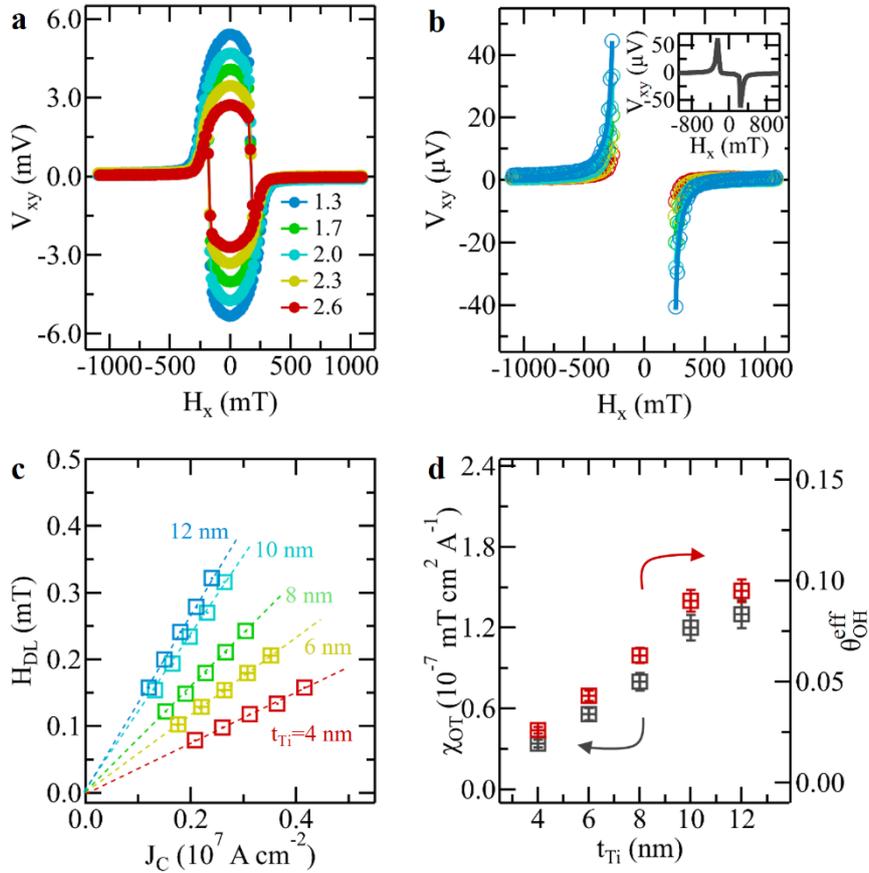

**Fig. 3. Determination of orbital torque efficiency. a,** First harmonic Hall voltage $V_{1\omega}$ as a function of the longitudinal magnetic field $H_x$ for Ti(10)/Fe$_{0.70}$Gd$_{0.30}$(10) at different current

densities. The unit of the current density is $10^6$ A cm$^{-2}$. **b,** The corresponding second harmonic Hall voltage $V_{2\omega}$ as a function of the longitudinal magnetic field $H_x$ at different current densities. **c,** The current density $J_C$ dependence of the damping-like effective field $H_{DL}$ for Ti($t_{Ti}$)/Fe$_{0.70}$Gd$_{0.30}$(10) with varying Ti thickness. **d,** The orbital torque efficiency $\chi_{OT}$ and effective orbital Hall angle $\theta_{OH}^{eff}$ as a function of Ti thickness $t_{Ti}$.

The effective orbital Hall angle $\theta_{OH}^{eff}$ can be estimated by using $\theta_{OH}^{eff} = 2e/\hbar \cdot M_s \cdot t_{FeGd} \cdot \chi_{OT}$, where e is the electron charge, $\hbar$ is the reduced Planck constant, $M_s$ (~163 emu/cc) is the saturation magnetization and $t_{FeGd}$ (~10 nm) is the ferrimagnetic layer thickness. As shown in Fig. 3d, the effective orbital Hall angle ($\theta_{OH}^{eff}$) have a similar trend with the orbital torque efficiency. The $\theta_{OH}^{eff}$ approaches ~0.1 when the Ti thickness is increased to 12 nm. This value is comparable to the reported effective spin Hall angle of traditional HMs like Pt, suggesting the OT to be an efficient way to switch perpendicular magnetization.

**Correlation between orbit torque and spin-orbit coupling**

A key feature of the OT is its correlation with SOC, as the orbital current is converted to the spin current via the SOC. Therefore, the effective orbital Hall angle ($\theta_{OH}^{eff} = \theta_{OH} \cdot \eta_{LS}$) is not only determined by the intrinsic orbital Hall angle ($\theta_{OH}$) of the LM, but also influenced by the orbital-to-spin conversion efficiency $\eta_{LS}$. Since $\eta_{LS}$ is related to the SOC of the magnetic layer, the OT can be probed and tuned through the chemical composition of the ferrimagnetic alloys in Ti/Fe$_{1-x}$Gd$_x$ magnetic bilayers. This is in stark contrast to the conventional SOT, where the role of the magnetic material choice is less important.

To confirm the SOC correlation of OT, the film stack of Ti(10)/Fe$_{0.80}$Gd$_{0.20}$(10) with less Gd content was also fabricated through the same growth condition. The Fe$_{0.80}$Gd$_{0.20}$ composition is magnetically Fe-dominated, which is contrast to the Gd-dominant Fe$_{0.70}$Gd$_{0.30}$. As shown in Fig. 4a, the polarity of the AHE loop for Fe$_{0.80}$Gd$_{0.20}$ is opposite to that of Fe$_{0.70}$Gd$_{0.30}$ (Fig. 2d). This is because the dominance of the spin-dependent transport property from the outer shell 3d magnetism of Fe element, as compared with the inner shell 4f magnetism of Gd element. So the anomalous Hall resistance mainly captures the magnetization direction of Fe.

As shown in the top panel of Fig. 4d, by fine tuning the magnetic composition, Fe$_{0.80}$Gd$_{0.20}$ and Fe$_{0.70}$Gd$_{0.30}$ possess nearly the same coercive field $H_C$, magnetic anisotropy field $H_k$, and saturation magnetization $M_S$ (Supplementary Figs. S1). In such a case, for OT switching, the main difference between these two systems is the strength of SOC due to the different Gd composition. In contrast, similar SOT efficiency is expected with these two different compositions, which has been observed in previous works [55-57].

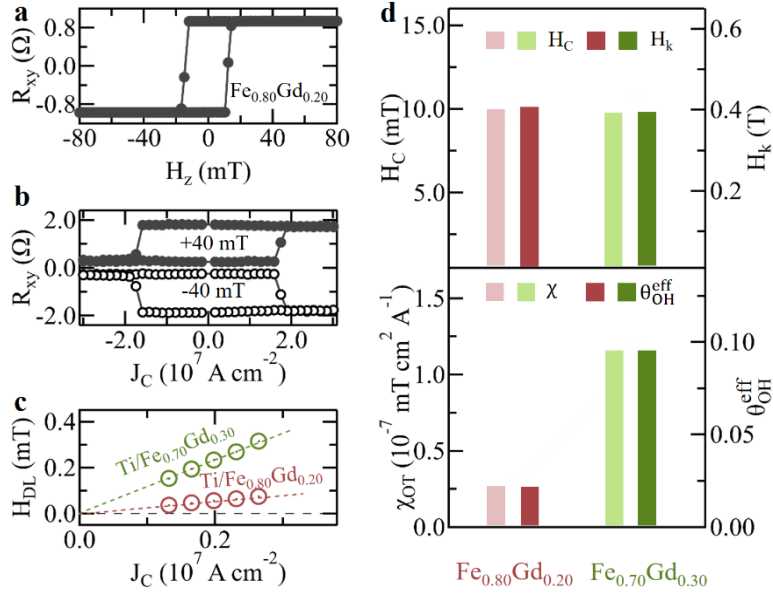

**Fig. 4. Correlation between orbit torque and spin-orbit coupling. a**, The anomalous Hall loops for Ti(10)/Fe$_{0.80}$Gd$_{0.20}$(10) sample. **b**, The corresponding current-induced magnetization switching curves under an in-plane field $H_x = \pm 40\ mT$. **c**, The current density $J_C$ dependence of the damping-like effective field $H_{DL}$ for Ti(10)/Fe$_{0.70}$Gd$_{0.30}$(10) and Ti(10)/Fe$_{0.80}$Gd$_{0.20}$(10) samples. **d**, The coercive field $H_C$ and magnetic anisotropy field $H_k$ for both samples in top panel and the corresponding orbital torque efficiency $\chi_{OT}$ and effective orbital Hall angle $\theta_{OH}^{eff}$ in bottom panel.

The current-induced switching and harmonic Hall measurements were then performed in the Ti(10)/Fe$_{0.80}$Gd$_{0.20}$(10) device. As shown in Fig. 4b, a deterministic magnetization switching is also achieved. The current density dependence of the effective damping-like field $H_{DL}$ for both Ti/Fe$_{0.80}$Gd$_{0.20}$ and Ti/Fe$_{0.70}$Gd$_{0.30}$ samples are plotted in Fig. 4c. As shown in the bottom panel of Fig. 4d, despite having nearly the same magnetic properties, the OT efficiency in Ti/Fe$_{0.70}$Gd$_{0.30}$ is increased by four-times comparing with that in Ti/Fe$_{0.80}$Gd$_{0.20}$. The critical switching current density for Fe$_{0.70}$Gd$_{0.30}$ is also smaller than that of Fe$_{0.80}$Gd$_{0.20}$ due to this larger OT efficiency. The enhancement of the OT efficiency is due to the larger orbital-to-spin conversion efficiency from to the increased Gd content and the corresponding larger SOC. Such strong correlation also confirms that the current-induced torque and the corresponding switching are dominated by the OT generated by the Ti light metal. Moreover, it also demonstrates that the OT efficiency can be systematically tuned through the chemical composition of ferrimagnet in LM/ferrimagnet bilayers.

**Conclusions**

We have successfully realized the orbital torque-induced switching of perpendicular ferrimagnet Fe$_{1-x}$Gd$_x$ by utilizing a typical light metal Ti. Due to the strong SOC of rare earth element Gd, an efficient orbital-to-spin current conversion can be directly achieved in Fe$_{1-x}$Gd$_x$ layer. The thickness dependent torque efficiency shows that the OT is dominated by the bulk effect of Ti. We found that the critical switching current density and orbital torque efficiency are comparable to the traditional heavy metals used for SOT. Furthermore, the correlation between OT and SOC of Fe$_{1-x}$Gd$_x$ is confirmed, making RE-TM ferrimagnetic material a versatile platform for investigating OT-related effects. Our work establishes the possibility of orbital torque-driven perpendicular magnetization reversal utilizing light metals in combination with RE-TM ferrimagnets, which is promising for developing spin-orbitronic devices.

**Methods**
**Sample growth and device fabrication**

Magnetic multilayers of stacking order Ta(1)/Ti($t_{Ti}$)/Fe$_{0.70}$Gd$_{0.30}$(10)/Si$_3$N$_4$(5), Ta(1)/Ti(10)/Fe$_{0.80}$Gd$_{0.20}$(10)/ Si$_3$N$_4$ (5), Si$_3$N$_4$ (5)/Fe$_{0.70}$Gd$_{0.30}$(10)/ Si$_3$N$_4$ (5) (number denotes the thickness in nanometer) are fabricated on thermally oxidized silicon substrates by using ultrahigh vacuum magnetron sputtering system (AJA Orion 8). The base pressure of the main chamber is better than $1\times10^{-8}$ Torr and the Ar pressure is at 3 mTorr. A 1 nm Ta layer is used as an adhesive layer and an upper insulating Si$_3$N$_4$ (3 nm) layer is used for preventing oxidization, respectively. The Fe$_{1-x}$Gd$_x$ FIM layers are synthesized by co-sputtering the Fe and Gd targets, in which the atomic ratio ($x$) can be adjusted by fixing the growth power of the Fe target while changing the growth power of the Gd target. Notably the deposition condition of Fe$_{1-x}$Gd$_x$ FIM layers are maintained the same for the Ti($t_{Ti}$)/Fe$_{0.70}$Gd$_{0.30}$(10)/ Si$_3$N$_4$ (5) multilayers with various Ti thickness. Magnetic multilayers were patterned into Hall bar devices with a channel width of 20 μm by utilizing standard photolithography and Ar ion milling.

**STEM imaging**

The interface and crystalline structure of the multilayers were examined using cross-sectional scanning transmission electron microscopy (STEM, Thermo Fischer Scientific Themis Z) operated at 300 kV. Thin section of the grown film are prepared using a focused ion beam (FIB) system for high resolution STEM examination.

**Magnetic property and electrical transport measurements**

The magnetic hysteresis loops were measured by using a magnetic property measurement system (MPMS, Quantum Design) with a maximum magnetic field of 7T. The magnetic fields were applied along the in-plane and out-of-plane direction at room temperature. The saturation magnetization and magnetic anisotropy field can be determined. The anomalous and planar Hall effect (AHE) measurements were performed by using a home-built electrical transport measurement system, in which the maximum magnetic field is 2T. The magnetic field is applied along the in-plane and out-of-plane direction. For all electrical transport measurements, standard Hall bar configuration was adopted and the electric current was set along the x axis.

**Current-induced switching and second harmonic measurements**

The current-induced magnetization switching is performed by injecting current pulses of duration 1 ms into the Hall bar devices and the corresponding AHE resistances were detected using a lock-in amplifier (SR830). A low-frequency alternating-current (ac) harmonic measurement using the lock-in technique is employed and two lock-in amplifiers are used to simultaneously record the first and second harmonic signals.

**Data availability**

The data supporting the findings of this study are available from the corresponding authors upon reasonable request.

**Acknowledgements**

This work was supported by the National Key R&D Program of China (Grant No. 2022YFA1403603), the National Natural Science Funds for Distinguished Young Scholar (52325105), the National Natural Science Foundation of China (12241406), the CAS Project for Young Scientists in Basic Research (YSBR-084), the Strategic Priority Research Program of the Chinese Academy of Sciences (Grant No. XDB33030100), the Chinese Academy of Sciences under contract No. JZHKYPT-2021-08.

**Author contributions**

T.X. and H.D. initiated the project and supervised the study. T.X., A.T. and K.W. grew the magnetic multilayers and fabricated the Hall bar devices. T.X. and A.T. characterized their electrical and magnetic properties. T.X. performed the current-induced switching and harmonic Hall measurements. T.X., Y.L. and H.D. wrote the manuscript with inputs from all authors. All authors discussed the results and commented on the manuscript.

**Competing interests**

The authors declare no competing interests.

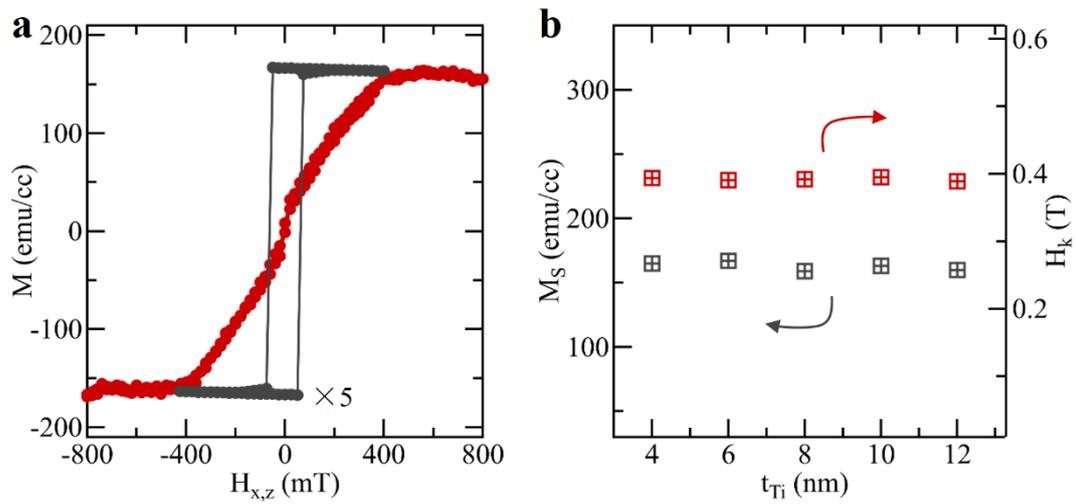

**Extended Data Fig. 1.** (a) The out-of-plane (black) and in-plane (red) magnetic hysteresis loops of Ti(10 nm)/Fe$_{0.70}$Gd$_{0.30}$/ Si$_3$N$_4$ sample. (b) The saturation magnetization M$_S$ and magnetic anisotropy field H$_k$ of multilayers with varying Ti thickness.

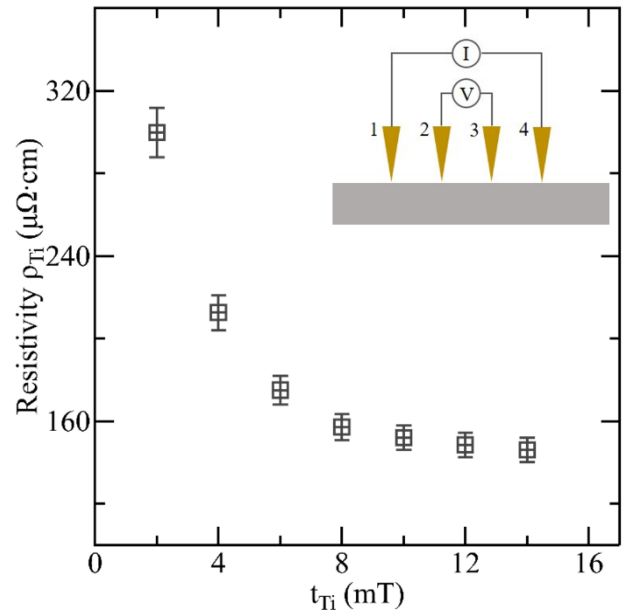

**Extended Data Fig. 2.** The thickness dependence of the electrical resistivity of Ti layer.

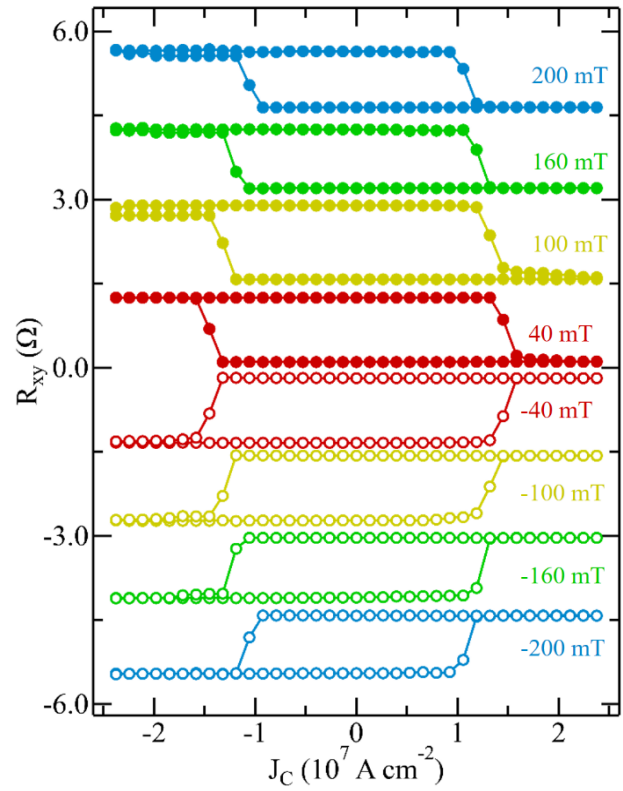

**Extended Data Fig. 3.** The current-induced switching measurements at different in-plane field $H_x$ for Ti(10 nm)/Fe$_{0.70}$Gd$_{0.30}$/ Si$_3$N$_4$ sample.

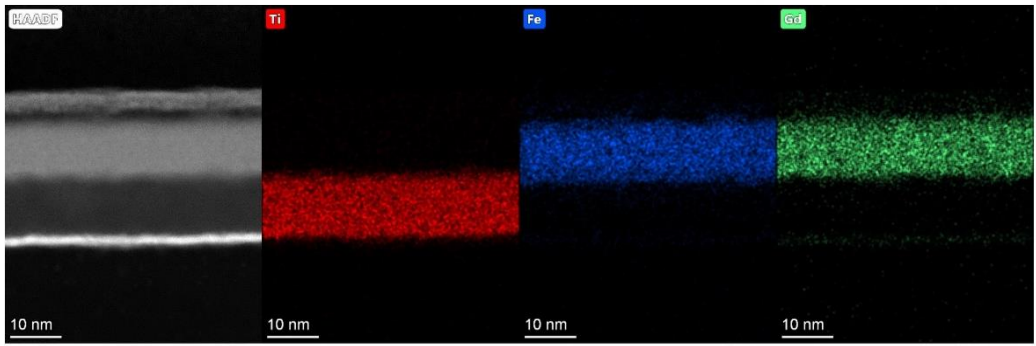

**Extended Data Fig. 4.** The cross-sectional STEM images of the Ti(10 nm)/Fe$_{0.70}$Gd$_{0.30}$/Si$_3$N$_4$ multilayer and the corresponding EDS element mapping for Ti, Fe, and Gd elements.

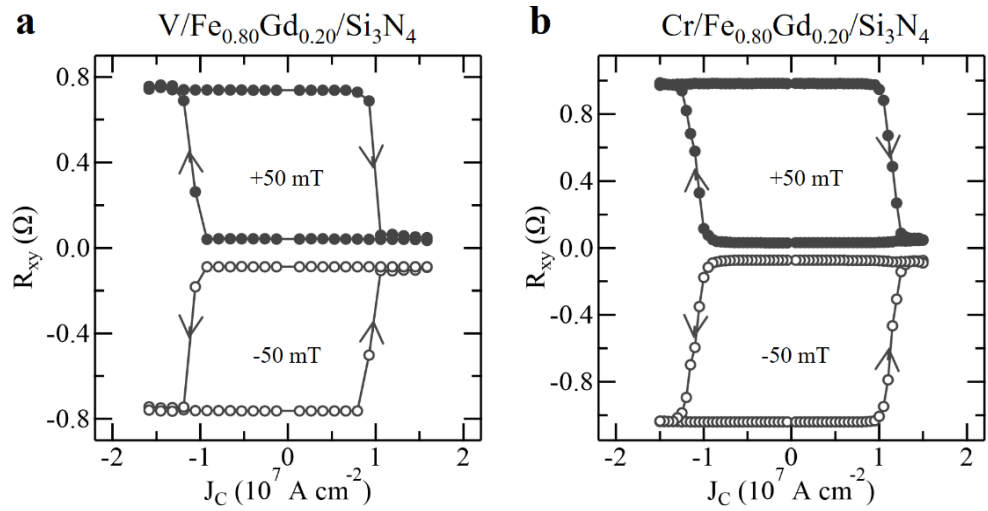

**Extended Data Fig. 5.** Current-induced orbital torque switching in V/Fe$_{0.80}$Gd$_{0.20}$/Si$_3$N$_4$ and Cr/Fe$_{0.80}$Gd$_{0.20}$/Si$_3$N$_4$ devices at $H_x = \pm 50\ mT$.

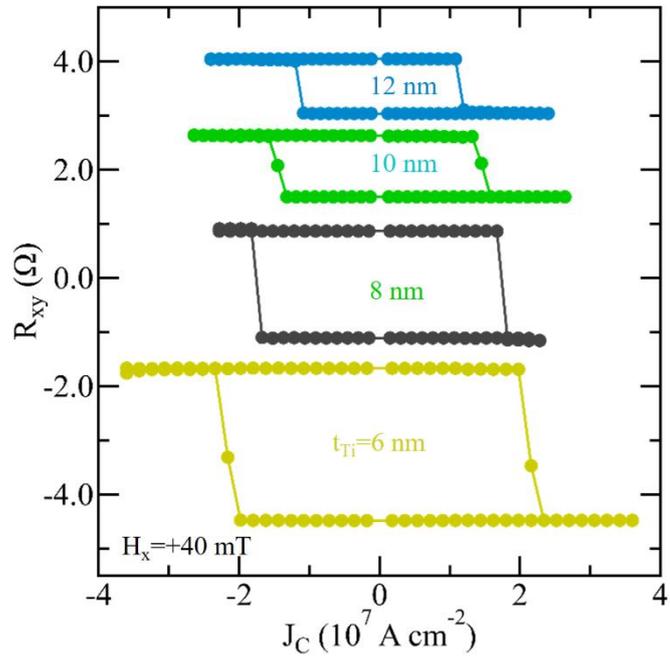

**Extended Data Fig. 6.** The current-induced switching measurements at $H_x = +40\ mT$ for Ti($t_{Ti}$)/Fe$_{0.70}$Gd$_{0.30}$/ Si$_3$N$_4$ samples with varying Ti thickness.